\documentclass[seceq]{ptptex}
\usepackage{graphicx} 
\usepackage{latexsym}
\notypesetlogo
\begin{document}
  \begin{flushright}
   RUP-09-01 \\
   April, 2009
\end{flushright}
\vspace{10mm} 

\begin{center}
\Large{   Drell-Yan  Cross Section in the Jet Calculus Scheme}

\vspace{15mm}

\large{Hidekazu {\sc Tanaka} and Hirokazu {\sc Kobayashi} \\
Department of Physics, Rikkyo University, Tokyo 171-8501, Japan     }
\vspace{25mm}

  {\Large ABSTRACT}
  \end{center}
        
\vspace{10mm}

We calculate factorized cross sections for lepton pair production mediated by a virtual photon in hadron-hadron collisions using the jet calculus scheme, in which a kinematical constraint due to parton radiation is taken into account.  This method guarantees a proper phase space boundary for subtraction terms.   Some properties of the calculated cross sections are examined.  We also discuss matching between the hard scattering cross sections and parton showers at the next-to-leading logarithmic (NLL) order of quantum chromodynamics (QCD). 


\section{Introduction}

In the evaluation of hadron-hadron scattering cross sections, logarithmic contributions due to collinear parton production are subtracted from the hard scattering cross section, and they are absorbed into the parton distributions of hadrons.   
However, the leading-logarithmic (LL) order of quantum chromodynamics (QCD) is insufficient to evaluate these processes.  Thus, the next-to-leading logarithmic (NLL)-order contributions should also be taken into account. The next-to-leading-order (NLO) calculation is also necessary in order to remove theoretical ambiguities due to the factorization procedure as well as the choice of the factorization scale. \cite{rf:1} 

In actual calculations, Monte Carlo methods are powerful tools for the evaluation of exclusive processes.   
 So far, parton shower models for initial-state parton radiation have been mainly constructed  on the basis of the LL order of QCD \cite{rf:2}, although some refinements, such as angular ordering conditions for soft gluon radiation, are taken into account. \cite{rf:3} 

 The scaling violation of parton distributions inside hadrons can be understood as the sequential evolution of partons in the initial state.  
 Conventionally, the scaling violation of the parton distributions is  calculated by solving the renormalization group equations in moments. Then these solutions are numerically inverted to yield the momentum fractions of partons.

 Alternatively, one can use parton shower models in order to evaluate the scaling violation of the parton distributions.  One such algorithm has been proposed in Ref. \citen{rf:4}, in which the parton showers are generated at the LL order of QCD using an algorithm consisting of a model based on the evolution of momentum distributions.  In this model, the scaling violation of the parton distributions is generated using only information from the splitting functions of the parton branching vertices and input distributions at a given energy.  It has been found that the method reproduces  the scaling violation of the flavor singlet parton distributions up to their normalizations.  This algorithm has been applied to the Drell-Yan process \cite{rf:5}, in which the NLO contributions for hard scattering cross sections have been included in a Monte Carlo model based on LL-order parton showers. 

  In order to allow the application of parton shower models to realistic processes, the matching problem should be solved, that is, double counting between hard scattering and parton showers must be avoided.
  This problem has been studied\cite{rf:6} in hadron-hadron collisions with  LL-order parton showers.  However, this method is insufficient, because the subtraction scheme dependence of the collinear singularity at the NLO cross section is not canceled by the factorization scheme dependence of the distribution functions.  

  It may be necessary to combine parton showers at the NLL order for the purpose of constructing more accurate Monte Carlo generators. As in the analytic calculations, it is desirable to combine NLO cross sections with parton showers at the NLL order of QCD. 
 
Recently, the parton shower model based on the evolution of momentum distributions was extended to the NLL order of QCD with the modified minimal subtraction ($\overline{\rm MS}$) scheme\cite{rf:1}\cite{rf:7}  as well as that with the jet calculus scheme.\cite{rf:8}

  In a previous paper,\cite{rf:9} we studied factorization schemes for the hard scattering cross sections in hadron-hadron collisions, in which the collinear singularities are subtracted using the $\overline{\rm MS}$ scheme.  In the conventional $\overline{\rm MS}$ scheme,  it has been pointed out that matching between the hard scattering cross section and the initial-state radiation is broken in the calculation of exclusive processes.  In order to implement factorization schemes that are appropriate for Monte Carlo methods using the parton shower models to the accuracy of the NLL order of QCD, we took into account a kinematical constraint due to parton radiation when considering the subtraction terms, which is called the $\overline{\rm MS}'$ scheme. The momenta of partons in the scattering processes are conserved in this method.  As an example, we have calculated  cross sections for the Drell-Yan lepton-pair production mediated by a virtual photon in hadron-hadron collisions.  We found that the conventional $\overline{\rm MS}$ scheme gives a double logarithmic term, which increases the cross section in the soft gluon limit, whereas the $\overline{\rm MS}'$ scheme gives only a single logarithmic term. The infrared behavior is reasonably stable in the case of factorization with the $\overline{\rm MS}'$ scheme.  However, in the collinear region after the collinear singularity is subtracted,  negative contributions remain for the cross section obtained with both the conventional $\overline{\rm MS}$ scheme and the $\overline{\rm MS}'$ scheme.  Such contributions cannot be ignored at the NLL-order accuracy. Event generation employed in the Monte Carlo methods with a negative probability may not be appropriate, since the strong cancellation between the negative contributions from the hard scattering cross section and the positive contributions from parton showers may give unstable results.

To solve this problem, we examine the jet calculus (${\rm JC}$) scheme studied in Ref. \citen{rf:8} for the calculation of the Drell-Yan process. In this scheme, an additional phase space factor of the hard scattering cross section is also subtracted. As shown in $\S$3, most  of the collinear contributions to the hard scattering cross section at the NLO are  subtracted in this scheme. 

  At the NLO accuracy, the processes $q{\bar q}\rightarrow \gamma^*g$, $qg\rightarrow \gamma^*q$ and ${\bar q}g\rightarrow \gamma^*{\bar q}$ contribute to the hard scattering cross section in actual physical systems.  In the following sections, we calculate factorized cross sections for the process  $q{\bar q}\rightarrow \gamma^*g$  using our method.  The factorized cross sections for ${\bar q}g\rightarrow \gamma^*{\bar q}$ are presented in the Appendix B.

 In $\S$2, we calculate factorization terms for the collinear singularity by the jet calculus scheme. Some properties of the factorized cross sections are presented in $\S$3.
Section 4 contains a summary and some comments.

\section{Factorization schemes}

In this section, we calculate factorization terms for the Drell-Yan lepton-pair production in quark $(q)$--antiquark $({\bar q})$ annihilation,
\begin{eqnarray}
 q(p_q)+{\bar q}(p_{\bar q}) \rightarrow \gamma^*(q)+g(p_g) \rightarrow \l^-(p_-)+l^+(p_+)+g(p_g), 
 \end{eqnarray}
mediated by a photon $\gamma^*$ with the virtuality $(p_- + p_+)^2=q^2=Q^2$, where a gluon $(g)$ is radiated in the final state. Here, $p_i~(i=q,{\bar q},g)$ and $p_{\pm}$ denote the momenta of the corresponding particles.

In the following calculation, the Mandelstam variables are defined by
\begin{eqnarray}
{\hat s}=(p_q+p_{\bar q})^2,~~{\hat t}=(p_q-p_g)^2, ~~{\hat u}=(p_{\bar q}-p_g)^2,
\end{eqnarray}
which satisfy ${\hat s}+{\hat t}+{\hat u}=Q^2$.

In order to obtain a finite cross section for the process $q{\bar q}\rightarrow \gamma^*g$, we subtract the collinear contributions due to the branching processes $q \rightarrow qg$ and ${\bar q} \rightarrow {\bar q}g$ from the hard scattering cross section.  Although these contributions  are compensated by the parton showers, the remnant of the subtracted cross section at the NLO cannot be ignored at the NLL-order accuracy.

In actual Monte Carlo simulations with parton shower models, the initial-state partons are evolved up to  a given energy scale $M$, which corresponds to  a factorization scale for the separation between the initial-state radiation and the hard scattering process.  The virtualities and longitudinal momentum fractions of the partons in the initial state are generated according to nonbranching probabilities (form factors) and the splitting functions of parton branching processes, respectively.\cite{rf:4}\cite{rf:7}\cite{rf:8}   The four-momenta of the partons are constructed from these values.  

The $s$-channel momentum ${\hat s}$ of the hard scattering process is constructed by the generated momenta  $p_q$ and $p_{\bar q}$ using Eq. ($2\cdot 2$). 
Then the virtuality of the photon $Q^2$  at the NLO is generated from the hard scattering cross section.
 
The momentum of the quark $r_q$, defined by $r_q=p_q-p_g$ for the branching process
\begin{eqnarray}
q(p_q) \rightarrow q(r_q)+g(p_g),
\end{eqnarray}
is described by in terms of the momentum fraction $z_q$, the virtuality $r_q^2$  and the transverse momentum $r_{qT}$, with $p_q\cdot r_{qT}=p_{\bar q}\cdot r_{qT}=0$, as 
\begin{eqnarray}
 r_q=p_q-p_g=z_qp_q+\alpha_q p_{\bar q}+r_{qT}, 
\end{eqnarray}
where $\alpha_q=r_q^2/{\hat s}$ for on-shell gluon radiation ($p_g^2=0$). 
Here, we set $p_q^2=p_{\bar q}^2=0$, because the relations $-p_q^2,-p_{\bar q}^2 \ll -r^2_q$ are expected in parton shower generation.

The squared momentum of a photon $Q^2$ is constructed as  
\begin{eqnarray}
Q^2=(r_q+p_{\bar q})^2
\end{eqnarray}
to $O(\alpha_s)$ accuracy of QCD.  Here, the quantity ${\hat \tau}=Q^2/{\hat s}$ is given by
\begin{eqnarray}
{\hat \tau}=z_q+\alpha_q.
\end{eqnarray}
 According to Eq. (2$\cdot$6), the phase space $0 \leq -r^2_q \leq M^2$ with $z_q \leq 1$ for the initial-state parton radiation  covers the region  $0 \leq -{\hat t} \leq (1-{\hat \tau}){\hat s}$ for $1-{\hat \tau}_M \leq {\hat \tau} \leq 1$ and $0 \leq -{\hat t} \leq M^2$ for ${\hat \tau} \leq 1-{\hat \tau}_M$ in the hard scattering cross section.  Here, $-{\hat t} = -r^2_q$ for $p_q^2=p_{\bar q}^2 =0$.  Therefore, we consider the two regions $1-{\hat \tau}_M \leq {\hat \tau} \leq 1$ and ${\hat \tau} \leq 1-{\hat \tau}_M$ separately. Here, we define ${\hat \tau}_M \equiv M^2/{\hat s}$. 

In the region $1-{\hat \tau}_M \leq {\hat \tau} \leq 1$, we subtract the collinear contributions from within the phase space to the hard scattering process, namely $0 \leq -{\hat t} \leq {\hat s}(1-{\hat \tau}) $.   
 If we ignore the term $\alpha_q=r^2_q/{\hat s}$ in Eq. (2$\cdot$6) and subtract the collinear contribution from the hard scattering cross section in the range $ 0 \leq -{\hat t} \leq M^2$, with ${\hat \tau}=z_q$, the matching between the hard scattering cross section and the initial state-radiation is broken.  In this case, $M^2$ is larger than the kinematical boundary for $-{\hat t}$, given by ${\hat s}(1-{\hat \tau})$ in the region $1-{\hat \tau}_M \leq {\hat \tau} \leq 1$.

As in the previous paper,\cite{rf:9}  we define, in $4-2\epsilon$ dimensions, the subtraction term divided by the Born cross section, ${\hat \sigma}_0(Q^2,\epsilon)$\footnote{The Born cross section ${\hat \sigma}_0(Q^2,\epsilon)$ is given in Ref. \citen{rf:9}.}, as
\begin{eqnarray}
 {dS_{qq}^{[F]} \over d{\hat \tau}d(-r^2_q)}=\int^1_0dz_q{d{\tilde S}_{qq}^{[F]} \over dz_q d(-r^2_q)}\delta(z_q-{\hat \tau}-(-r^2_q)/{\hat s})
 \end{eqnarray}
with
 \begin{eqnarray}
 {d{\tilde S}_{qq}^{[F]} \over dz_q d(-r^2_q)}={\alpha_s \over 2\pi}{1 \over \Gamma(1-\epsilon)}\left[{-r^2_q \over 4\pi\mu^2}\right]^{-\epsilon}{P_{qq}^{[F]}(z_q,\epsilon) \over -r^2_q},
 \end{eqnarray}
  where the coefficient of the subtraction term  $P_{qq}^{[F]}(z_q,\epsilon)$ depends on the factorization scheme $F$. In $4-2\epsilon$ dimensions, the strong coupling constant is defined by $\alpha_s\mu^{2\epsilon}$ for the dimensionless coupling $\alpha_s$ and a mass parameter $\mu$.

We define the integrated contribution as 
\begin{eqnarray}
 \int_0^{{\hat s}w^{[F](I)}} d(-r^2_q){dS_{qq}^{[F]} \over d{\hat \tau}d(-r^2_q)}= {\alpha_s \over 2\pi}{1 \over \Gamma(1-\epsilon)}\left[{{\hat s} \over 4\pi\mu^2}\right]^{-\epsilon}{\tilde F}_{qq}^{[F](I)}(\epsilon,{\hat \tau}),
 \end{eqnarray}
 where $I$ denotes the region of phase space for the hard scattering process being considered. Here, ${\hat s}w^{[F](I)}$ is a limit of the $-r^2_q$ integration.
In our previous paper\cite{rf:9}, we implemented the $\overline{\rm MS}$ scheme, in which 
 \begin{eqnarray}
  P_{qq}^{[\overline{\rm MS}]}(z_q,\epsilon)&=&\left({\hat P}_{qq}^{(0)}(z_q)\right)_+,
   \end{eqnarray}
  for the branching process presented in Eq. (2$\cdot$3),
 with
\begin{eqnarray}
  {\hat P}_{qq}^{(0)}(z_q)=C_F{1+z_q^2 \over 1-z_q}. 
\end{eqnarray}
 Here, $C_F=4/3$ is the color factor, and we define 
\begin{eqnarray}
 \left({\hat f}({\hat \tau})\right)_+={\hat f}({\hat \tau})-\delta(1-{\hat \tau})\int^1_0dy{\hat f}(y) 
\end{eqnarray}
for a function ${\hat f}({\hat \tau})$ unregulated at ${\hat \tau}=1$. 

In this paper, we examine the ${\rm JC}$ scheme\cite{rf:8},  in which  the coefficient of the subtraction term is given by 
\begin{eqnarray}
  P_{qq}^{[{\rm JC}]}(z_q,\epsilon)=\left({\hat P}_{qq}^{[{\rm JC}]}(z_q,\epsilon)\right)_++\delta(1-z_q){\epsilon \over 2}C_F,
   \end{eqnarray}
where
\begin{eqnarray}
{\hat P}_{qq}^{[{\rm JC}]}(z_q,\epsilon)= (1-z_q)^{-\epsilon}{\hat P}_{qq}(z_q,\epsilon) \simeq {\hat P}_{qq}^{(0)}(z_q)-\epsilon {\hat Q}_{qq}^{[\rm JC]}(z_q)
\end{eqnarray}
for $\epsilon \ll 1$, with
\begin{eqnarray}
{\hat Q}_{qq}^{[\rm JC]}(z_q)={\hat P}^{(0)}_{qq}(z_q)\log(1-z_q)+C_F(1-z_q).
\end{eqnarray}
 Here,\cite{rf:10}
 \begin{eqnarray}
  {\hat P}_{qq}(z_q,\epsilon)={\hat P}_{qq}^{(0)}(z_q)+\epsilon{\hat P}'_{qq}(z_q) 
 \end{eqnarray}
with
\begin{eqnarray}
  {\hat P}'_{qq}(z_q)=-C_F(1-z_q).
 \end{eqnarray}
 In this scheme, the factor $(1-z_q)^{-\epsilon}$ from the phase space, which gives a negative contribution at the NLO, and the $O(\epsilon)$ term of the splitting function ${\hat P}'_{qq}(z_q)$ in $4-2\epsilon$ dimensions for the branching process\cite{rf:10}  are also subtracted from the hard scattering cross section.\cite{rf:8} \footnote{The function $Q_{qq}^{[\rm JC]}(z)$ corresponds to $\Delta F_{qq}^{[{\rm JC}]}(z) $ in Ref. \citen{rf:8}} As shown in $\S$3, most  of the collinear contributions to the hard scattering cross section at the NLO are  subtracted in this scheme. In Eq. (2$\cdot$13), the regularization of the infrared singularity is defined so as to conserve the total momentum of the initial-state partons.

First, we consider the region $1-{\hat \tau}_M \leq {\hat \tau} \leq 1$ ($I=S$).
For $z_q={\hat \tau}(F={\rm JC})$ and $w^{[F](S)}={\hat \tau}_M$, the subtraction term  for the soft  gluon radiation is given by  
\begin{eqnarray}
  & & {\tilde F}^{[{\rm JC}](S)}_{qq}(\epsilon,{\hat \tau}) =  \int_0^{M^2} d(-r^2_q)\left[{-r^2_q \over {\hat s}}\right]^{-\epsilon}{P_{qq}^{[\rm JC]}({\hat \tau},\epsilon) \over -r^2_q} \nonumber \\
 & & ~~ =C_F\Big[\left({1\over -\epsilon}+\log{\hat \tau}_M\right){1 \over C_F}\left({\hat P}_{qq}^{(0)}({\hat \tau})\right)_++(1+{\hat \tau}^2)\left({\log(1-{\hat \tau}) \over 1-{\hat \tau}}\right)_+ \nonumber \\
& & ~~~~~ +1-{\hat \tau}-{11 \over 4}\delta(1-{\hat \tau})\Big].
\end{eqnarray}
Similarly, with $z_q={\hat \tau}+(-r^2_q)/{\hat s}$ ($F={\rm JC}'$) and $w^{[{\rm JC}'](S)}=1-{\hat \tau}$, we obtain 
\begin{eqnarray}
  & & {\tilde F}^{[{\rm JC}'](S)}_{qq}(\epsilon,{\hat \tau}) =  \int_0^{(1-{\hat \tau}){\hat s}} d(-r^2_q)\left[{-r^2_q \over {\hat s}}\right]^{-\epsilon}{P_{qq}^{[\rm JC]}({\hat \tau}+(-r^2_q)/{\hat s},\epsilon) \over -r^2_q} \nonumber \\
 & & ~~ = C_F\Big[{1\over -\epsilon}{1 \over C_F}\left({\hat P}_{qq}^{(0)}({\hat \tau})\right)_++2(2+{\hat \tau}^2)\left({\log(1-{\hat \tau}) \over 1-{\hat \tau}}\right)_+  \nonumber \\
& & ~~~~~ +{3 \over 2}{1 \over (1-{\hat \tau})_+}-\left({1 \over 3}\pi^2+{11 \over 4}\right)\delta(1-{\hat \tau})\Big].
\end{eqnarray}
 The calculation method employed in this region is explained in Ref. \citen{rf:9}.

The factorization term in the region ${\hat \tau}_0 \leq {\hat \tau} < 1-{\hat \tau}_M~(I=C)$ is given by  
\begin{eqnarray}
 & & {\tilde F}^{[{\rm JC}](C)}_{qq}(\epsilon,{\hat \tau}) =  \int_0^{M^2} d(-r^2_q)\left[{-r^2_q \over {\hat s}}\right]^{-\epsilon}{{\hat P}_{qq}^{[\rm JC]}({\hat \tau},\epsilon) \over -r^2_q} \nonumber \\
& & ~~= C_F\left[{1 \over C_F} {\hat P}^{(0)}_{qq}({\hat \tau})\left({1\over -\epsilon}+\log{\hat \tau}_M(1-{\hat \tau})\right)+1-{\hat \tau}\right]
\end{eqnarray}
for the ${\rm JC}$ scheme.  Here, we define ${\hat \tau}_0=Q^2_0/{\hat s}$, where $Q_0$ is a lower limit for the virtuality of the photon to be observed.

For the ${\rm JC}'$ scheme, we obtain
\begin{eqnarray}
  & & {\tilde F}^{[{\rm JC}'](C)}_{qq}(\epsilon,{\hat \tau}) =  \int_0^{M^2} d(-r^2_q)\left[{-r^2_q \over {\hat s}}\right]^{-\epsilon}{{\hat P}_{qq}^{[\rm JC]}({\hat \tau}+(-r^2_q)/{\hat s},\epsilon) \over -r^2_q} \nonumber \\
& & ~~ =C_F\Big[{1 \over C_F}  {\hat P}_{qq}^{(0)}({\hat \tau})\left({1 \over -\epsilon}+\log{(1-{\hat \tau})^2{\hat \tau}_M \over 1-{\hat \tau}-{\hat \tau}_M}\right) +(1+{\hat \tau})\log{1-{\hat \tau} \over 1-{\hat \tau}-{\hat \tau}_M} \nonumber \\
& & ~~~~ +1-{\hat \tau}-{\hat \tau}_M \Big].
\end{eqnarray}

The calculation in the collinear gluon region is given in Appendix A.

  By replacing $q$ with ${\bar q}$ in the above equations, we obtain the subtraction terms for the antiquark legs, which are the same as those for the quark legs.

\section{Factorized cross sections}

First, we calculate the contributions of the soft gluon radiation, which are important for evaluating the NLO contribution for the Born cross section, namely ${\hat s}\simeq Q^2$. 
  The factorized cross section to the soft gluon region $1-{\hat \tau}_M \leq {\hat \tau} \leq 1$ ($I=S$) in the factorization scheme $F$ is defined by
\begin{eqnarray}
{d{\hat \sigma}_{q{\bar q}}^{[F](S)} \over d{\hat \tau}}=
{\alpha_s \over 2\pi}{\hat \sigma}_0(Q^2,0)K_{q{\bar q}}^{[F](S)}({\hat \tau})
\end{eqnarray}
with 
\begin{eqnarray}
K_{q{\bar q}}^{[F](S)}({\hat \tau})=K_{q{\bar q}}({\hat \tau},\epsilon)-{\tilde F}^{[F](S)}_{qq}(\epsilon,{\hat \tau})-{\tilde F}^{[F](S)}_{{\bar q}{\bar q}}(\epsilon,{\hat \tau})
\end{eqnarray}
and
\begin{eqnarray}
{\hat \sigma}_0(Q^2,0)={4\pi\alpha^2 \over 3N_CQ^2}{\hat e}_q^2.
\end{eqnarray}
 Here, $N_C=3$ and the electric coupling constant of the quark is defined by ${\hat e}_q^2\alpha$ for $-\epsilon \rightarrow 0$.

The differential cross section for gluon radiation  with virtual loop contributions is given by \cite{rf:9}\cite{rf:11}\footnote{Our common factor in the hard scattering cross section and the factorization term is different from that defined in Ref. \citen{rf:11}. Therefore, no ${\hat P}_{qq}^{(0)}({\hat \tau})\log{\hat \tau}$ term appears in Eq. (3$\cdot$5). }
\begin{eqnarray}
{d{\hat \sigma}_{q{\bar q}} \over d{\hat \tau}}={\alpha_s \over 2\pi}{\hat \sigma}_0(Q^2,\epsilon){\Gamma(1-\epsilon) \over \Gamma(1-2\epsilon)}\left[{{\hat s} \over 4\pi\mu^2}\right]^{-\epsilon}K_{q{\bar q}}({\hat \tau},\epsilon)
\end{eqnarray}
with
\begin{eqnarray}
K_{q{\bar q}}({\hat \tau},\epsilon)&=&C_F\Big[{2 \over -\epsilon}{1 \over C_F}\left({\hat P}_{qq}^{(0)}({\hat \tau})\right)_++4(1+{\hat \tau}^2)\left({\log(1-{\hat \tau}) \over 1-{\hat \tau}}\right)_+ \nonumber \\
 & & + \left(-8+{2 \over 3}\pi^2\right)\delta(1-{\hat \tau})\Big].
\end{eqnarray}

From Eq. (3$\cdot$2), we obtain 
\begin{eqnarray}
K_{q{\bar q}}^{[{\rm JC}](S)}({\hat \tau})&=&C_F\Big[2(1+{\hat \tau}^2)\left({\log(1-{\hat \tau}) \over 1-{\hat \tau}}\right)_+ -2(1-{\hat \tau})\nonumber \\
& & - {2 \over C_F}\left({\hat P}_{qq}^{(0)}({\hat \tau})\right)_+\log{\hat \tau}_M + \left(-{5 \over 2}+{2 \over 3}\pi^2\right)\delta(1-{\hat \tau})\Big]
\end{eqnarray}
with the ${\rm JC}$ scheme and 
\begin{eqnarray}
K_{q{\bar q}}^{[{\rm JC}'](S)}({\hat \tau})&=&C_F\Big[-4\left({\log(1-{\hat \tau}) \over 1-{\hat \tau}}\right)_+-{3 \over (1-{\hat \tau})_+} \nonumber \\
& & + \left(-{5 \over 2}+{4 \over 3}\pi^2\right)\delta(1-{\hat \tau})\Big]
\end{eqnarray}
with the ${\rm JC}'$ scheme.

The factorized cross section in the ${\rm JC}$ scheme depends on the factorization scale $M$. With the ${\rm JC}'$ scheme, the factorized cross section has no $M$ dependence, because the subtraction term is integrated over the range $0 \leq -r_q^2 \leq {\hat s}(1-{\hat \tau})$. In this region, the virtuality of the photon $Q^2$ is constructed from the momenta of the partons generated by parton showers, in accordance with Eq. (2$\cdot$6),  within the region $-r^2_q \leq M^2$ for $z_q \leq 1$ . Therefore, the factorization scale for the distribution functions in this region is also given  by $M^2$.

In order to evaluate the soft gluon contribution, we integrate Eq. (3$\cdot$1) over the range $(1-\eta_s){\hat s} \leq Q^2 \leq {\hat s}$, with fixed ${\hat s}$, where ${\hat s}$ is generated by the parton showers. Here, $\eta_s$ is a cutoff parameter satisfying $\eta_s \leq {\hat \tau}_M$. 
 The energy scale of the running coupling constant in this region can be chosen as $Q^2$. In order to simplify our analysis, we evaluate the soft gluon contribution with the coupling constant $\alpha_s(Q^2)\simeq\alpha_s({\hat s})$.  The running coupling constant for the accuracy of the NLL order is normalized as $\alpha_s(M_Z^2)=0.114$ at the $Z^0$ boson mass.\cite{rf:12}

The integrated cross section is given by 
\begin{eqnarray}
\sigma_{q{\bar q}}^{[F](S)} (1,1-\eta_s)\simeq{\alpha_s({\hat s}) \over 2\pi}{\hat \sigma}_0({\hat s},0)I_{q{\bar q}}^{[F](S)} (1,1-\eta_s)
\end{eqnarray}
with
\begin{eqnarray}
I_{q{\bar q}}^{[F](S)} (1,1-\eta_s)=\int^1_{1-\eta_s} {d{\hat \tau} \over {\hat \tau}}K^{[F](S)}_{q{\bar q}}({\hat \tau}).
\end{eqnarray}

The calculated results for the range $1-\eta_s \leq {\hat \tau} \leq 1$ are given by 
\begin{eqnarray}
I_{q{\bar q}}^{[{\rm JC}](S)}(1,1&-&\eta_s)= C_F\Big[2\left\{SP_-(\eta_s,0)-\eta_s(\log\eta_s-2)+\log^2\eta_s+\log(1-\eta_s)\right\} \nonumber \\
& & -2\left\{-\log(1-\eta_s)-\eta_s+{3 \over 2}+2\log\eta_s\right\}\log{\hat \tau}_M-{5 \over 2}+{2 \over 3}\pi^2\Big] 
\end{eqnarray}
with the ${\rm JC}$ scheme, and 
\begin{eqnarray}
I_{q{\bar q}}^{[{\rm JC}'](S)}(1,1&-&\eta_s)=  C_F\Big[-4SP_-(\eta_s,0)-2\log^2\eta_s +3\log{1-\eta_s \over \eta_s} \nonumber \\
& & -{5 \over 2}+{4 \over 3}\pi^2\Big] 
\end{eqnarray}
with the ${\rm JC}'$ scheme.  Here, we define 
\begin{eqnarray}
SP_{-}(a,b) \equiv \int^a_bd{\hat \tau}{\log{\hat \tau} \over 1 - {\hat \tau}}.
\end{eqnarray}

We calculate the $\eta_s$ dependence of the function 
\begin{eqnarray}
R^{[F](S)}(\eta_s)=1+{\sigma_{q{\bar q}}^{[F](S)} (1,1-\eta_s) \over {\hat \sigma}_0({\hat s},0)}\simeq1+{\alpha_s({\hat s}) \over 2\pi}I_{q{\bar q}}^{[F](S)} (1,1-\eta_s)
\end{eqnarray}
using  Eqs. (3$\cdot$10) and (3$\cdot$11).

In Fig. 1, the $\eta_s$ dependence of the contribution obtained with the ${\rm JC}'$ scheme is plotted by the solid curve. The result obtained with the ${\rm JC}$ scheme for $\eta_s={\hat \tau}_M \leq 0.5$, which corresponds to the contribution integrated over the range $1-{\hat \tau}_M \leq {\hat \tau} \leq 1$, is represented by the dash-dotted curve.   The $\eta_s$ dependences obtained from the two schemes become similar, because part of the term $\log\eta_s\log{\hat \tau}_M$  is canceled by the $\log^2\eta_s$ term in the ${\rm JC}$ scheme, represented by Eq. (3$\cdot$10).  

 The dashed curve represents the $\eta_s$ dependence obtained with the ${\rm JC}$ scheme at ${\hat \tau}_M=0.5$.  The function $I_{q{\bar q}}^{[{\rm JC}](S)} $ near the threshold ($\eta_s \ll {\hat \tau}_M$) increases as $I_{q{\bar q}}^{[{\rm JC}](S)}\sim2C_F\log^2\eta_s$, whereas the results obtained with the ${\rm JC}$ scheme for $\eta_s={\hat \tau}_M$ and with the ${\rm JC}'$ scheme decrease as $I_{q{\bar q}}^{[F](S)}\sim -2C_F\log^2\eta_s$ in the soft gluon region.   

 The error in the calculation is approximated by $1 \leq \alpha_s((1-\eta_s){\hat s})/\alpha_s({\hat s}) < 1.05$ for $0 \leq \eta_s \leq {\hat \tau}_M=0.5$ at $\sqrt{\hat s}=200~{\rm GeV}$.

As shown in Fig. 1, the soft gluon contribution depends on the kinematical boundary for the subtraction term.  The result integrated over the region $1-{\hat \tau}_M \leq {\hat \tau} \leq 1$ with the ${\rm JC}$ scheme  behaves similarly to  that with the ${\rm JC}'$  scheme. Therefore, these two schemes give similar results in the soft gluon region.  
\begin{figure}
\centerline{\includegraphics[width=10cm]{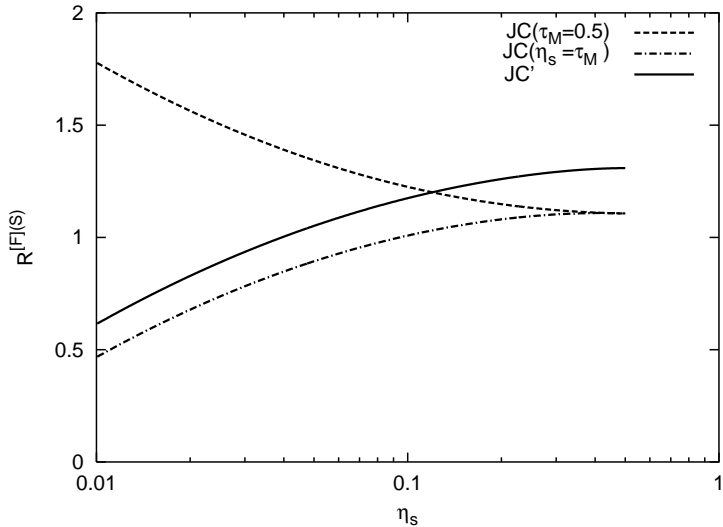}}
\caption{The integrated contributions of the cross sections divided by ${\hat \sigma}_0({\hat s},0)$, which are defined by Eq. (3$\cdot$13) with $\sqrt{\hat s}=200~{\rm GeV}$.  The solid curve represents the result obtained with the ${\rm JC}'$ scheme.  With the ${\rm JC}$ scheme, the dashed curve and dash-dotted curve represent the results when ${\hat \tau}_M=0.5$ and ${\hat \tau}_M=\eta_s$, respectively.  
}
\end{figure}

Next, we calculate the remnant in the region ${\hat \tau}_0 \leq {\hat \tau} < 1-{\hat \tau}_M~(I=C)$, which is given by
\begin{eqnarray}
 K_{q{\bar q}}^{[F](C)}({\hat \tau})&=&K_{q{\bar q}}^{(C)}({\hat \tau},\epsilon)-{\tilde F}^{[F](C)}_{qq}(\epsilon,{\hat \tau}).
\end{eqnarray}
The collinear contribution of the hard scattering cross section for the real gluon radiation integrated over the range $0 \leq -{\hat t} \leq M^2$ ( or  $0 \leq -{\hat u} \leq M^2$) is given by 
\begin{eqnarray}
\int^{M^2}_0d(-{\hat t}){d{\hat \sigma}^{(r)}_{q{\bar q}} \over d{\hat \tau}d(-{\hat t})}={\alpha_s \over 2\pi}{\hat \sigma}_0(Q^2,\epsilon){1 \over \Gamma(1-\epsilon) }\left[{{\hat s} \over 4\pi\mu^2}\right]^{-\epsilon}K_{q{\bar q}}^{(C)}({\hat \tau},\epsilon)
\end{eqnarray}
with
\begin{eqnarray}
K_{q{\bar q}}^{(C)}({\hat \tau},\epsilon)= C_F\Big[{1 \over C_F}{\hat P}_{qq}^{(0)}({\hat \tau})\left({1 \over -\epsilon}+   \log{{\hat \tau}_M(1-{\hat \tau})^2 \over 1-{\hat \tau}-{\hat \tau}_M}\right)+1-{\hat \tau}-2{\hat \tau}_M\Big],
\end{eqnarray}
where the differential cross section $d{\hat \sigma}^{(r)}_{q{\bar q}}/d{\hat \tau}/d(-{\hat t})$ is given in Ref. \citen{rf:9}.
The calculation of the remnant in the hard collinear region is presented in Appendix A.
 We obtain 
\begin{eqnarray}
 K_{q{\bar q}}^{[{\rm JC}](C)}({\hat \tau})&=&C_F\left[{1 \over C_F}{\hat P}_{qq}^{(0)}({\hat \tau})\log{(1-{\hat \tau}) \over 1-{\hat \tau}-{\hat \tau}_M}-2{\hat \tau}_M \right]
\end{eqnarray}
with the ${\rm JC}$ scheme and
\begin{eqnarray}
K_{q{\bar q}}^{[{\rm JC}'](C)}({\hat \tau})&=&  C_F\Big[-(1+{\hat \tau})\log{(1-{\hat \tau}) \over 1-{\hat \tau}-{\hat \tau}_M} -{\hat \tau}_M\Big]
\end{eqnarray}
with the ${\rm JC}'$ scheme.

Here, we calculate the ${\hat \tau}$ dependence of the function 
\begin{eqnarray}
R^{[F](C)}({\hat \tau})={1 \over {\hat \sigma}_0({\hat s},0)}{d\sigma_{q{\bar q}}^{[F](C)} \over  d{\hat \tau} }= {\alpha_s(Q^2) \over 2\pi{\hat \tau}}K_{q{\bar q}}^{[F](C)} ({\hat \tau})
\end{eqnarray}
using Eqs. (3$\cdot$17) and (3$\cdot$18). 

In Fig. 2, the ${\hat \tau}$ dependence of the contribution obtained with the ${\rm JC}'$ scheme is plotted by the solid curve.   The result obtained with the ${\rm JC}$ scheme is represented by the dashed curve. 
 Most of the collinear contribution is subtracted from the cross section in the ${\rm JC}'$ scheme, whereas a positive contribution remains near ${\hat \tau} \sim 1$ for the result obtained with the ${\rm JC}$ scheme. 

 We also show the calculated results obtained from the $\overline{\rm MS}$ and $\overline{\rm MS}'$ schemes by the dotted curves. These schemes were studied in our previous paper. \cite{rf:9} Explicit expressions for these schemes are given in Appendix A.  

In Fig. 2, the hard scattering contribution integrated over $M^2 \leq -{\hat t} \leq {\hat s}(1-{\hat \tau})-M^2$ is represented by the dash-dotted curve.

As shown in Fig. 2, the remnant of the hard collinear contribution for the ${\rm JC}'$ scheme can be safely neglected even at NLL-order accuracy. We can avoid the large cancellation between the hard scattering cross section and the parton showers. Therefore, we expect reasonably stable results in Monte Carlo calculations with this scheme. On the other hand, the negative contributions of the NLO cross sections obtained with the $\overline{\rm MS}$ and $\overline{\rm MS}'$ schemes should be canceled by the contributions from the parton showers.

\begin{figure}
\centerline{\includegraphics[width=10cm]{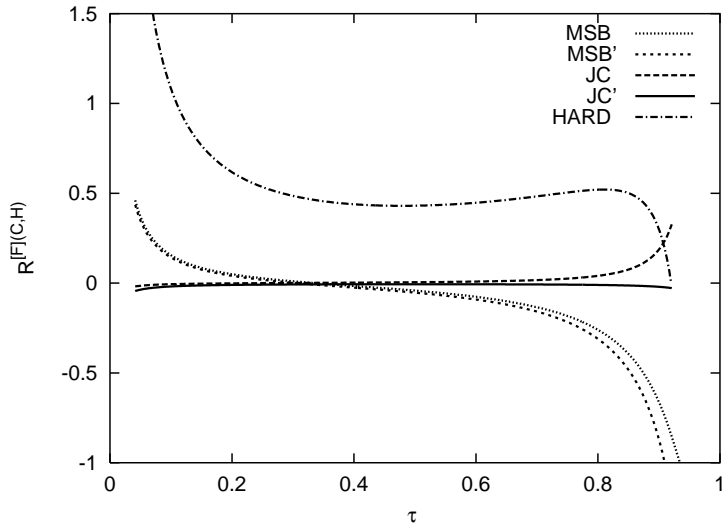}}
\caption{The ${\hat \tau}$ dependence of the factorized cross sections with $\sqrt{\hat s}=10^3$ ${\rm GeV}$ and $M=Q_0=200$ ${\rm GeV}$ for the process $q{\bar q} \rightarrow \gamma^*g$. The solid curve represents the result obtained with the ${\rm JC}'$ scheme. The result obtained with the ${\rm JC}$ scheme is represented by the dashed curve. The calculated results obtained from the $\overline{\rm MS}$ and $\overline{\rm MS}'$ schemes are shown by the dotted curves. The hard scattering contribution integrated over $M^2 \leq -{\hat t} \leq {\hat s}(1-{\hat \tau})-M^2$ is represented by the dash-dotted curve.}
\end{figure}
Although the parton showers already include the collinear contributions of the processes $g \rightarrow qX$ and $g \rightarrow {\bar q}X$, as well as  those of  $q \rightarrow qX$ and ${\bar q} \rightarrow {\bar q}X$, at NLL order, we have to add the NLO contributions for the processes $qg \rightarrow \gamma^*q$ and ${\bar q}g \rightarrow \gamma^*{\bar q}$ in the hard scattering region. 
The calculation of the factorized cross sections is rather straightforward. We present the NLO contributions for the process ${\bar q}g \rightarrow \gamma^*{\bar q}$ in Appendix B.  The numerical results of the factorized cross sections for this process are plotted in Fig. 3, in which the notation is the same as that in Fig. 2.  In this case, the remnant of the hard collinear contribution with the ${\rm JC}'$ scheme can also be neglected at NLL-order accuracy.

\begin{figure}
\centerline{\includegraphics[width=10cm]{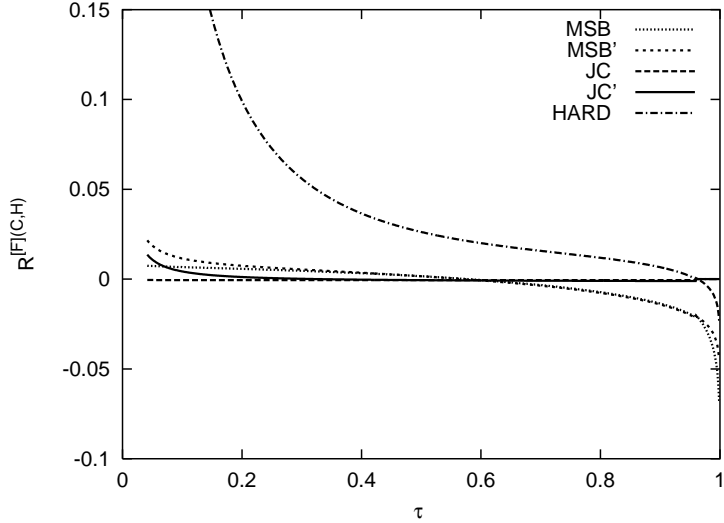}}
\caption{The ${\hat \tau}$ dependence of the factorized cross sections with $\sqrt{\hat s}=10^3$ ${\rm GeV}$ and $M=Q_0=200$ ${\rm GeV}$ for the process ${\bar q}g \rightarrow \gamma^*{\bar q}$ ( $qg \rightarrow \gamma^*q$ ). The notation is the same as that in Fig. 2.}
\end{figure}

\section{Summary and comments}

In this paper, we have studied collinear factorization for hard scattering cross sections in hadron-hadron collisions, in which the initial-state radiation is generated by parton shower models. 

We calculated hard scattering cross sections for the process $q{\bar q} \rightarrow \gamma^*g$ with the jet calculus scheme, in which a kinematical constraint for the gluon radiation is taken into account. This method is called the ${\rm JC}'$ scheme.  In this method, this kinematical constraint guarantees a proper phase space boundary for the subtraction terms.  
We also calculated the hard scattering cross section for this process without using the kinematical constraint (${\rm JC}$ scheme). In the ${\rm JC}$ scheme, matching between the hard scattering cross sections and the initial-state radiation is broken for $1-{\hat \tau}_M \leq {\hat \tau} \leq 1$.

The subtracted cross section integrated over the range $1-\eta_s \leq {\hat \tau} \leq 1$ obtained with the ${\rm JC}$ scheme increases as $\sim 2C_F\log^2\eta_s$  in the soft gluon region ($\eta_s \ll {\hat \tau}_M$), whereas the results obtained with the ${\rm JC}$ scheme for $\eta_s={\hat \tau}_M$ and with the ${\rm JC}'$ scheme decrease as $\sim -2C_F\log^2\eta_s$.  
The kinematical constraint due to the parton radiation changes the properties of the NLO cross sections in the soft gluon region.
If the soft gluon contributions are resummed as 
\begin{eqnarray}
{\hat \sigma}_0({\hat s},0)+\sigma_{q{\bar q}}^{[{\rm JC}'](S)} (1,1-\eta_s)\simeq{\hat \sigma}_0({\hat s},0)\exp\left[{\alpha_s({\hat s}) \over 2\pi}I_{q{\bar q}}^{[{\rm JC}'](S)} (1,1-\eta_s)\right],
\end{eqnarray}  
the cross sections obtained with the ${\rm JC}'$ scheme and with the ${\rm JC}$ scheme for $\eta_s={\hat \tau}_M$ vanish at the soft gluon limit, which is a desirable property for Monte Carlo methods. 

In the hard collinear region ${\hat \tau}_0 \leq {\hat \tau} < 1-{\hat \tau}_M$,  after the collinear singularity is subtracted, negative contributions remain for ${\hat \tau} \sim 1$ for the cross section obtained with both the $\overline{\rm MS}$ scheme and the $\overline{\rm MS}'$ scheme.  Such contributions  may not be appropriate for event generation employed in the Monte Carlo methods with NLL-order accuracy, because large cancellation may occur between the hard scattering contributions and those from the initial-state radiation, whereas most of the collinear contributions are subtracted from the cross section obtained with the ${\rm JC}'$ scheme.  Therefore, with the ${\rm JC}'$ scheme, we expect to obtain reasonably stable results in the numerical calculations.

Finally we comment on the factorization scheme dependence between the initial parton radiation generated by the parton showers at the NLL order and the hard scattering cross sections at the NLO.  The parton evolution depends on the subtraction scheme used for the mass singularity, whose appearance is due to collinear parton production. 
 
 In parton shower models, infrared unregulated splitting functions are used for evaluation of the nonbranching probabilities.   For gluon radiation from a quark leg, the difference between the splitting function obtained with the $\overline{\rm MS}$ scheme and that obtained with the jet calculus scheme is given by  $\beta_0{\hat Q}_{qq}^{[\rm JC]}(z_q)/2 \simeq \beta_0{\hat Q}_{qq}^{[\rm JC]}({\hat \tau})/2$ at NLL-order accuracy.\cite{rf:1} Here, $\beta_0=11-2/3N_f$ with $N_f$ flavour quarks.

In order to cancel the factorization scheme dependence, we use  the relation
  \begin{eqnarray}
 {\hat P}_{ij}^{[{\rm JC}](1)}(z)={\hat P}_{ij}^{[\overline{\rm MS}](1)}(z)-{\beta_0 \over 2}{\hat Q}_{ij}^{[{\rm JC}]}(z)
\end{eqnarray}
with
\begin{eqnarray}
{\hat Q}_{ij}^{[\rm JC]}(z)={\hat P}^{(0)}_{ij}(z)\log(1-z)-{\hat P}'_{ij}(z)
\end{eqnarray}
for $i,j=q,{\bar q},g$ at the NLL order in order to obtain nonbranching probabilities \cite{rf:8}.  Here, $P_{ij}^{[\overline{\rm MS}](1)}(z)$ and $P_{ij}^{[{\rm JC}](1)}(z)$ are the splitting functions calculated with the $\overline{\rm MS}$ scheme\cite{rf:13} and ${\rm JC}$ scheme\cite{rf:8}, respectively. The functions ${\hat P}'_{ij}(z)$ are the $O(\epsilon)$ terms of the splitting functions in $4-2\epsilon$ dimensions.\cite{rf:10}. 
 In this case, the soft gluon contribution of the splitting functions may yield large higher-order contributions for $z \sim 1$. 
In particular, the functions ${\hat Q}_{qq}^{[{\rm JC}]}(z)$ and ${\hat Q}_{gg}^{[{\rm JC}]}(z)$ behave as $ \sim \log(1-z)/(1-z) $ for $z \rightarrow 1$. 
  Such terms can be absorbed into the strong coupling constant\cite{rf:8} as 
\begin{eqnarray}
{\alpha_s\left(K^2\right) \over 2\pi}{\hat P}^{(0)}_{ij}(z)   & +&  \left({\alpha_s\left(K^2\right) \over 2\pi}\right)^2{\hat P}^{{[\rm JC}](1)}_{ij}(z)  \nonumber \\
& & \simeq  {\alpha_s\left({\bar K}^2_{ij}\right) \over 2\pi}{\hat P}^{(0)}_{ij}(z)   +  \left({\alpha_s\left({\bar K}^2_{ij}\right) \over 2\pi}\right)^2{\hat P}^{[\overline{\rm MS}](1)}_{ij}(z) 
\end{eqnarray}
with ${\bar K}_{ij}^2=e^{{\hat P}'_{ij}(z)/{\hat P}^{(0)}_{ij}(z)}(1-z)K^2 \simeq (0.4 \sim 1)k_T^2$,
where $k_T$ is the transverse momentum of the generated parton with spacelike virtuality $-K^2 < 0$. 

The jet calculus  scheme presented in this paper may be useful for the construction of a more accurate Monte Carlo algorithm, in which the parton radiation at the NLL order as well as the hard scattering cross sections at the NLO of QCD are taken into account. 

 In future work, we shall construct a Monte Carlo algorithm with the ${\rm JC}'$ scheme using the method presented in this paper.

\section*{Acknowledgements}

This work was supported in part by the Rikkyo University Special Fund for Research.


\newpage
\begin{center}
{\Large Appendix A}
\end{center}
\vspace{5mm}

In this appendix, we derive the remnant in the hard collinear region ($I=C$) for the process $q{\bar q}\rightarrow \gamma^*g$ . 
 
The factorization terms are given by  
\begin{eqnarray*}
 {\tilde F}^{[F](C)}_{qq}(\epsilon,{\hat \tau})= \left({1\over -\epsilon}+\log{\hat \tau}_M\right) {\hat P}_{qq}^{[F]}({\hat \tau},\epsilon)
\end{eqnarray*}
for $F=\overline{\rm MS},{\rm JC}$, where 
\begin{eqnarray*}
  {\hat P}_{qq}^{[\overline{\rm MS}]}({\hat \tau},\epsilon)={\hat P}^{(0)}_{qq}({\hat \tau})
\end{eqnarray*}
and 
\begin{eqnarray*}
  {\hat P}_{qq}^{[{\rm JC}]}({\hat \tau},\epsilon)={\hat P}^{(0)}_{qq}({\hat \tau})-\epsilon{\hat Q}_{qq}^{[\rm JC]}({\hat \tau}).
\end{eqnarray*}
The functions ${\hat P}^{(0)}_{qq}$ and ${\hat Q}_{qq}^{[\rm JC]}$ are given by Eqs. (2$\cdot$11) and (2$\cdot$15) in the main text, respectively.

For $\overline{\rm MS}'$ and ${\rm JC}'$, we obtain
\begin{eqnarray*}
 {\tilde F}^{[\overline{\rm MS}'](C)}_{qq}(\epsilon,{\hat \tau}) &=&C_F\Big[{1 \over C_F}  {\hat P}_{qq}^{(0)}({\hat \tau})\left({1 \over -\epsilon}+\log{(1-{\hat \tau}){\hat \tau}_M \over 1-{\hat \tau}-{\hat \tau}_M}\right) \nonumber \\
& & +(1+{\hat \tau})\log{1-{\hat \tau} \over 1-{\hat \tau}-{\hat \tau}_M}-{\hat \tau}_M \Big]
\end{eqnarray*}
and
\begin{eqnarray*}
 {\tilde F}^{[{\rm JC}'](C)}_{qq}(\epsilon,{\hat \tau}) &=&{\tilde F}^{[\overline{\rm MS}'](C)}_{qq}(\epsilon,{\hat \tau})+{\hat Q}_{qq}^{[\rm JC]}({\hat \tau}),
\end{eqnarray*}
respectively.

The remnant in the hard collinear region is defined by
\begin{eqnarray*}
 K_{q{\bar q}}^{[F](C)}({\hat \tau})&=&K_{q{\bar q}}^{(C)}({\hat \tau},\epsilon)-{\tilde F}^{[F](C)}_{qq}(\epsilon,{\hat \tau}),
\end{eqnarray*}
where $K_{q{\bar q}}^{(C)}({\hat \tau},\epsilon)$ is given by  Eq. (3$\cdot$16) in the main text.
We obtain  
\begin{eqnarray*}
 K_{q{\bar q}}^{[\overline{\rm MS}](C)}({\hat \tau})&=& C_F\left[{1 \over C_F}{\hat P}_{qq}^{(0)}({\hat \tau})\log{(1-{\hat \tau})^2 \over 1-{\hat \tau}-{\hat \tau}_M}+(1-{\hat \tau}-2{\hat \tau}_M) \right]
\end{eqnarray*}
with the $\overline{\rm MS}$ scheme and
\begin{eqnarray*}
K_{q{\bar q}}^{[\overline{\rm MS}'](C)}({\hat \tau})&=&  K_{q{\bar q}}^{[\overline{\rm MS}](C)}({\hat \tau})+C_F\left[{2 \over 1-{\hat \tau}}\log{1-{\hat \tau}-{\hat \tau}_M \over 1-{\hat \tau}}+{\hat \tau}_M\right]
\end{eqnarray*}
with the $\overline{\rm MS}'$ scheme.  The results obtained with the ${\rm JC}$ scheme and  ${\rm JC}'$ scheme are respectively given by
\begin{eqnarray*}
 K_{q{\bar q}}^{[{\rm JC}](C)}({\hat \tau})&=& K_{q{\bar q}}^{[\overline{\rm MS}](C)}({\hat \tau})-{\hat Q}_{qq}^{[{\rm JC}]}({\hat \tau})
\end{eqnarray*}
and
\begin{eqnarray*}
K_{q{\bar q}}^{[{\rm JC}'](C)}({\hat \tau})&=&   K_{q{\bar q}}^{[\overline{\rm MS}'](C)}({\hat \tau})-{\hat Q}_{qq}^{[{\rm JC}]}({\hat \tau}),
\end{eqnarray*}
which are presented in Eqs. (3$\cdot$17) and (3$\cdot$18) in  the main text.

The contribution of the hard gluon radiation ($I=H$) integrated over the range $M^2 \leq -{\hat t} \leq {\hat s}(1-{\hat \tau})-M^2$ is given by 
\begin{eqnarray*}
 K^{(H)}_{q{\bar q}}({\hat \tau})&=&{1 \over {\hat s}}\int^{{\hat s}(1-{\hat \tau})-M^2}_{M^2}d(-{\hat t})C_F\left[{\hat s}\left({1 \over -{\hat t}}+{1 \over -{\hat u}}\right){1 \over C_F} {\hat P}_{qq}^{(0)}({\hat \tau})-2\right] \nonumber \\
&=& 2C_F\left[{1 \over C_F} {\hat P}_{qq}^{(0)}({\hat \tau})\log{1-{\hat \tau}-{\hat \tau}_M \over {\hat \tau}_M}-(1-{\hat \tau}-2{\hat \tau}_M)\right] 
\end{eqnarray*}
for $ {\hat \tau} \leq 1-2{\hat \tau}_M$.

\vspace{5mm}

\begin{center}
{\Large Appendix B}
\end{center}
\vspace{5mm}

In this appendix, we calculate the cross section of the Drell-Yan lepton-pair production in antiquark $({\bar q})$--gluon $(g)$ scattering,
\begin{eqnarray*}
 {\bar q}(p_{\bar q})+g(p_g) \rightarrow \gamma^*(q)+{\bar q}(p'_{\bar q}) \rightarrow \l^-(p_-)+l^+(p_+)+{\bar q}(p'_{\bar q}), 
 \end{eqnarray*}
mediated by a photon $\gamma^*$ with virtuality $(p_- + p_+)^2=q^2=Q^2$. 
Here, $p_i~(i={\bar q},g)$ and $p_{\pm}$ denote the momenta of the corresponding particles. 

The momentum of the quark $r_q$, defined by $r_q=p_g-p_q$ for the branching process $g(p_g) \rightarrow q(r_q)+{\bar q}(p'_{\bar q})$, is described by using the momentum fraction $z_q$, the virtuality $r_q^2$ and the transverse momentum $r_{qT}$, with $p_{\bar q}\cdot r_{qT}=p_g\cdot r_{qT}=0$, as 
\begin{eqnarray*}
 r_q=z_qp_g+\alpha_q p_{\bar q}+r_{qT}, 
\end{eqnarray*}
where $\alpha_q=r_q^2/{\hat s}$ for ${p'}_{\bar q}^2=0$ with $p_g^2=0$.

In the $\overline{\rm MS}$ scheme, the coefficient of the subtraction term is given by 
 \begin{eqnarray*}
  {\hat P}_{qg}^{[\overline{\rm MS}]}(z_q,\epsilon)={\hat P}_{qg}^{(0)}(z_q)
 \end{eqnarray*}
 with
\begin{eqnarray*}
  {\hat P}_{qg}^{(0)}(z_q)=T_R[z_q^2+(1-z_q)^2]. 
\end{eqnarray*}
 Here, $T_R=1/2$ is the color factor.

In  the jet calculus scheme,  we define\cite{rf:8}
\begin{eqnarray*}
  {\hat P}_{qg}^{[{\rm JC}]}(z_q,\epsilon)={\hat P}_{qg}^{(0)}(z_q)-\epsilon {\hat Q}_{qg}^{[{\rm JC}]}(z_q)
\end{eqnarray*}
with
\begin{eqnarray*}
{\hat Q}_{qg}^{[\rm JC]}(z_q)={\hat P}^{(0)}_{qg}(z_q)\log(1-z_q)+z_q(1-z_q).
\end{eqnarray*}

For $F=\overline{\rm MS},{\rm JC}$,  the subtraction terms  for  $1-{\hat \tau}_M \leq {\hat \tau} \leq 1~(I=S)$ and ${\hat \tau}_0 \leq {\hat \tau} < 1-{\hat \tau}_M~(I=C)$ are given by  
\begin{eqnarray*}
 {\tilde F}^{[F](I)}_{qg}(\epsilon,{\hat \tau}) =  \left({1\over -\epsilon}+\log{\hat \tau}_M\right){\hat P}_{qg}^{[F]}({\hat \tau},\epsilon).
\end{eqnarray*}
For $F=\overline{\rm MS}',{\rm JC}'$, we obtain 
\begin{eqnarray*}
 {\tilde F}^{[\overline{\rm MS}'](S)}_{qg}(\epsilon,{\hat \tau}) &=& T_R\Big[{1\over -\epsilon}{1 \over T_R}{\hat P}_{qg}^{(0)}({\hat \tau})+{1 \over T_R}{\hat P}_{qg}^{(0)}({\hat \tau})\log(1-{\hat \tau})  \nonumber \\
& & + (1-{\hat \tau})(3{\hat \tau}-1)\Big]
\end{eqnarray*}
and
\begin{eqnarray*}
 {\tilde F}^{[{\rm JC}'](S)}_{qg}(\epsilon,{\hat \tau}) &=&{\tilde F}^{[\overline{\rm MS}'](S)}_{qg}(\epsilon,{\hat \tau})+{\hat Q}_{qg}^{[\rm JC]}({\hat \tau})
\end{eqnarray*}
 for $1-{\hat \tau}_M \leq {\hat \tau} \leq 1$ ($I=S$), and  
\begin{eqnarray*}
 {\tilde F}^{[\overline{\rm MS}'](C)}_{qg}(\epsilon,{\hat \tau}) &=& T_R\Big[{1\over -\epsilon}{1 \over T_R}{\hat P}_{qg}^{(0)}({\hat \tau})+{1 \over T_R}{\hat P}_{qg}^{(0)}({\hat \tau})\log{\hat \tau}_M  \nonumber \\
& & + 2(2{\hat \tau}-1){\hat \tau}_M+{\hat \tau}_M^2\Big]
\end{eqnarray*}
and
\begin{eqnarray*}
 {\tilde F}^{[{\rm JC}'](C)}_{qg}(\epsilon,{\hat \tau}) ={\tilde F}^{[\overline{\rm MS}'](C)}_{qg}(\epsilon,{\hat \tau})+{\hat Q}_{qg}^{[\rm JC]}({\hat \tau})
\end{eqnarray*}
for the region ${\hat \tau}_0 \leq {\hat \tau} < 1-{\hat \tau}_M~(I=C)$, respectively.

We obtain the subtracted contributions  
\begin{eqnarray*}
 K_{{\bar q}g}^{[F](I)}({\hat \tau})=K_{{\bar q}g}^{(I)}({\hat \tau},\epsilon)-{\tilde F}^{[F](I)}_{qg}(\epsilon,{\hat \tau})
\end{eqnarray*}
with
\begin{eqnarray*}
& & K_{{\bar q}g}^{(S)}({\hat \tau},\epsilon)={1 \over {\hat s}}\int^{(1-{\hat \tau}){\hat s}}_0d(-{\hat t})\left[{(-{\hat t})(-{\hat u}) \over {\hat s}^2}\right]^{-\epsilon}{\tilde K}_{{\bar q}g}^{(r)}({\hat \tau},-{\hat t},\epsilon)  \nonumber \\
& & ~~~~~=T_R\Big[{1 \over T_R}{\hat P}_{qg}^{(0)}({\hat \tau})\left({1 \over -\epsilon}+   \log(1-{\hat \tau})^2-1\right)+{3 \over 2}+{\hat \tau}-{3 \over 2}{\hat \tau}^2\Big]
\end{eqnarray*}
for $I=S$, and 
\begin{eqnarray*}
& & K_{{\bar q}g}^{(C)}({\hat \tau},\epsilon)={1 \over {\hat s}}\int^{M^2}_0d(-{\hat t})\left[{(-{\hat t})(-{\hat u}) \over {\hat s}^2}\right]^{-\epsilon}{\tilde K}_{qg}^{(r)}({\hat \tau},-{\hat t},\epsilon)  \nonumber \\
& & ~~~~~=T_R\Big[{1 \over T_R}{\hat P}_{qg}^{(0)}({\hat \tau})\left({1 \over -\epsilon}+   \log{\hat \tau}_M(1-{\hat \tau})\right)+2{\hat \tau}(1-{\hat \tau}+{\hat \tau}_M)+{1 \over 2}{\hat \tau}_M^2\Big]
\end{eqnarray*}
for $I=C$.  The differential cross section for the process ${\bar q}g \rightarrow {\bar q}\gamma^*$ is given by\cite{rf:11}
 \begin{eqnarray*}
{d{\hat \sigma}^{(r)}_{{\bar q}g} \over d{\hat \tau}d(-{\hat t})}={\alpha_s \over 2\pi{\hat s}} {\hat \sigma}_0(Q^2,\epsilon)\left[{(-{\hat t})(-{\hat u}) \over 4\pi{\hat s}\mu^2}\right]^{-\epsilon}{1 \over \Gamma(1-\epsilon)}{\tilde K}_{{\bar q}g}^{(r)}({\hat \tau},-{\hat t},\epsilon)
\end{eqnarray*}
 with
 \begin{eqnarray*}
 {\tilde K}_{{\bar q}g}^{(r)}({\hat \tau},-{\hat t},\epsilon)=T_R\left[{{\hat s} \over -{\hat t}}{1 \over T_R}{\hat P}_{qg}({\hat \tau},\epsilon)+2{\hat \tau}+{-{\hat t} \over {\hat s}}\right], 
\end{eqnarray*}
where   
\begin{eqnarray*}
  {\hat P}_{qg}({\hat \tau},\epsilon)={\hat P}_{qg}^{(0)}({\hat \tau})+\epsilon {\hat P}'_{qg}({\hat \tau})
 \end{eqnarray*}
 with
\begin{eqnarray*}
  {\hat P}'_{qg}({\hat \tau})=-{\hat \tau}(1-{\hat \tau}). 
\end{eqnarray*}
Here, the Born cross section ${\hat \sigma}_0(Q^2,\epsilon)$ is given in Ref. \citen{rf:9}.

  For $I=S$, we obtain
\begin{eqnarray*}
 K_{{\bar q}g}^{[\overline{\rm MS}](S)}({\hat \tau})= T_R\left[{1 \over T_R}{\hat P}_{qg}^{(0)}({\hat \tau})\left(\log{(1-{\hat \tau})^2 \over {\hat \tau}_M}-1\right)+{3 \over 2}+{\hat \tau}-{3 \over 2}{\hat \tau}^2 \right]
\end{eqnarray*}
with the $\overline{\rm MS}$ scheme and
\begin{eqnarray*}
K_{{\bar q}g}^{[\overline{\rm MS}'](S)}({\hat \tau})= T_R\left[{1 \over T_R}{\hat P}_{qg}^{(0)}({\hat \tau})\left(\log(1-{\hat \tau})-1\right)+{5 \over 2}-3{\hat \tau}+{3 \over 2}{\hat \tau}^2 \right]
\end{eqnarray*}
with the $\overline{\rm MS}'$ scheme.

  For $I=C$, we have
\begin{eqnarray*}
 K_{{\bar q}g}^{[\overline{\rm MS}](C)}({\hat \tau})= T_R\left[{1 \over T_R}{\hat P}_{qg}^{(0)}({\hat \tau})\log(1-{\hat \tau})+2{\hat \tau}(1-{\hat \tau}+{\hat \tau}_M)+{1 \over 2}{\hat \tau}_M^2 \right]
\end{eqnarray*}
and
\begin{eqnarray*}
K_{{\bar q}g}^{[\overline{\rm MS}'](C)}({\hat \tau})= T_R\left[{1 \over T_R}{\hat P}_{qg}^{(0)}({\hat \tau})\log(1-{\hat \tau})+2{\hat \tau}(1-{\hat \tau}-{\hat \tau}_M)+{1 \over 2}{\hat \tau}_M(4-{\hat \tau}_M) \right].
\end{eqnarray*}

The calculated results with the ${\rm JC}$ scheme and ${\rm JC}'$ scheme are respectively given by
\begin{eqnarray*}
 K_{{\bar q}g}^{[{\rm JC}](I)}({\hat \tau})= K_{{\bar q}g}^{[\overline{\rm MS}](I)}({\hat \tau})-{\hat Q}_{qg}^{[{\rm JC}]}({\hat \tau})
\end{eqnarray*}
and
\begin{eqnarray*}
K_{{\bar q}g}^{[{\rm JC}'](I)}({\hat \tau})=  K_{{\bar q}g}^{[\overline{\rm MS}'](I)}({\hat \tau})-{\hat Q}_{qg}^{[{\rm JC}]}({\hat \tau}),
\end{eqnarray*}
for $I=S,C$. The numerical results are presented in the main text.

For ${\hat \tau} \rightarrow 1$, the functions $K_{{\bar q}g}^{[F](S)}({\hat \tau})$ behave as 
\begin{eqnarray*}
 K_{{\bar q}g}^{[\overline{\rm MS}](S)}({\hat \tau}) \rightarrow T_R\log{(1-{\hat \tau})^2 \over {\hat \tau}_M}, ~~~
K_{{\bar q}g}^{[\overline{\rm MS}'](S)}({\hat \tau}) \rightarrow T_R\log(1-{\hat \tau})
\end{eqnarray*}
for $F=\overline{\rm MS},\overline{\rm MS}'$, and  
\begin{eqnarray*}
K_{{\bar q}g}^{[{\rm JC}](S)}({\hat \tau}) \rightarrow T_R\log{(1-{\hat \tau}) \over {\hat \tau}_M}, ~~~
K_{{\bar q}g}^{[{\rm JC}'](S)}({\hat \tau}) \rightarrow 0
\end{eqnarray*}
for $F={\rm JC},{\rm JC}'$. 
The integrated results for the range $1-\eta_s \leq {\hat \tau} \leq 1$, 
\begin{eqnarray*}
I_{{\bar q}g}^{[F](S)} (1,1-\eta_s)=\int^1_{1-\eta_s} {d{\hat \tau} \over {\hat \tau}}K^{[F](S)}_{{\bar q}g}({\hat \tau}),
\end{eqnarray*}
vanish for $\eta_s \rightarrow 0$.

The contribution of the hard gluon radiation ($I=H$) integrated over the range $M^2 \leq -{\hat t} \leq {\hat s}(1-{\hat \tau})$ is given by 
\begin{eqnarray*}
 K^{(H)}_{{\bar q}g}({\hat \tau})&=&{1 \over {\hat s}}\int^{{\hat s}(1-{\hat \tau})}_{M^2}d(-{\hat t}){\tilde K}_{{\bar q}g}^{(r)}({\hat \tau},-{\hat t},0)  \nonumber \\
&=&  T_R\left[{1 \over T_R}{\hat P}_{qg}^{(0)}({\hat \tau})\log{(1-{\hat \tau}) \over {\hat \tau}_M}+2{\hat \tau}(1-{\hat \tau}-{\hat \tau}_M)+{1 \over 2}\{(1-{\hat \tau})^2-{\hat \tau}_M^2\}\right]
\end{eqnarray*}
for $ {\hat \tau} \leq 1-{\hat \tau}_M$.  

 The subtracted cross sections for the quark legs  are the same as those for the antiquark legs.

\end{document}